\documentstyle[12pt,twoside,epsf, epsfig, fleqn,esprc1]{article}

\newcommand{\AmS}{{\protect\the\textfont2
  A\kern-.1667em\lower.5ex\hbox{M}\kern-.125emS}}

\title{Precise strength of the $\pi$NN coupling constant}

\author{T. E. O. Ericson\thanks{This work was partly done at the Research 
Center for the Subatomic Structure of Matter in Adelaide.}$^,$\address{The 
Svedberg Laboratory, Box 533, S-75121 Uppsala, Sweden}$^,$\address{CERN, 
CH-1211 Geneva 23, Switzerland},
B. Loiseau\address{Universit\'e P. \& M. Curie, L.P.T.P.E., F-75252 Paris,
France},
J. Rahm$^a$, J.  Blomgren$^a$,  N.  Olsson$^a$ and 
A.~W.~Thomas\address{Special Research Center for the Subatomic Structure 
of Matter,\\ University of Adelaide, Adelaide 5005, Australia}}

\begin{document}
\maketitle

\begin{abstract} Abstract.  We report here a preliminary value for the
$\pi N\! N$
coupling constant deduced from the GMO sumrule for forward $\pi N$
scattering. As in our previous determination from np backward differential
scattering cross sections we give a critical discussion
of the analysis with careful attention not only to the statistical,
 but
also to the systematic uncertainties.  Our preliminary
evaluation gives $g^2_c(GMO)=13.99(24)$.  
\end{abstract}

\section {INTRODUCTION}

 The crucial coupling of low energy hadron physics is the $\pi N\! N$
coupling constant, which for the pseudoscalar interaction of a
charged pion has the approximate value $g^2_c \simeq 14$.  One would like
this quantity to be determined experimentally to a precision of about 1\%
for accurate tests of chiral symmetry predictions, such as the
Goldberger-Treiman relation.  Determinations of the coupling constant in
later years are given in Table 1.

 \begin{table}[h] \caption[h]{Some
important determinations of  the pion-nucleon coupling constant}

\begin{tabular}{lcclll}
    Source                        &
   Year             &System &   $g^2_{\pi N\! N}$&Reference    & \\
\hline
 Karlsruhe-Helsinki
 &   1980 &$\pi$p
  &    $14.28 (18)$&    Nucl. Phys. {\bf A336}, 331 (1980).&    \\
Kroll et al. 
&1981&pp&14.52(40)&Physics Data {\bf
22-1}(1981).\\
\hline
Nijmegen~
\cite{Sto93a}
&  1993& pp, np
&   $13.58(5)$   &Phys. Rev. C {\bf  47}, 512 (1993).
 & \\
VPI
&  1994& pp, np
&   13.7   &Phys. Rev. C {\bf 50}, 2731 (1994).&      \\
Nijmegen 
&1997 &pp, np& 13.54(5)&$\Pi $N
Newsletter {\bf 13}, 96 (1997).& \\
Timmermans
&1997&$\pi^+$p&13.45(14) &$\Pi $N Newsletter {\bf 13}, 80 (1997).& \\
\hline
VPI \cite {Arn94b}
  &1994   &GMO,  $\pi$p
&  $13.75 (15)$   & Phys. Rev. C {\bf 49}, 2729 (1994).&   \\
Uppsala~\cite{Eri95} &  1995 &np$\rightarrow $pn &
$14.62(30)$ &Phys. Rev. Lett.
{\bf 75}, 1046 (1995).& \\
Uppsala \cite{Rah98}&1998&np$\rightarrow $pn&14.52(26)&Phys. Rev. C {\bf
57}, 1077 (1998).&
\\ \hline
\end{tabular} \label{tab:coupling constants} \end{table} 

The Nijmegen
group pointed out some years ago that the earlier determinations from the
1980's had important systematic uncertainties and they have since
advocated values about 5\% lower than the previous ones, mainly based on
their analysis of $N\! N$ interactions \cite{Sto93a}.  However, these
later
determinations 
are, in general, not transparently linked to the underlying data and the
systematic errors in the analysis are unknown. An exception is the GMO
analysis by Arndt et al. \cite {Arn94b}. Important physical constants are
generally determined directly from experimental data with transparent,
refutable procedures.  The $\pi N\! N$ coupling constant should be no
exception.  We have therefore started a program of such
determinations \cite {{Eri95},{Rah98}}. A first approach is based
on single energy
backward np differential cross sections, dominated by pion pole
contributions.  The extrapolation  to the pion pole at
$t=-{\bf q}^2= m^2_{\pi}$ gives directly $g^4$.  This is based on an old
idea of
Chew, which has not been workable in practice for the following
reasons: 
1)  previous data were not precise enough and in particular lacked
absolute normalization, 
2) the original extrapolation method requires
a polynomial expansion with a large number of terms, which makes
systematics in the extrapolation obscure.  

These deficiencies have been largely
eliminated \cite {{Eri95},{Rah98}}. High precision absolutely
normalized
differential np cross sections have recently been measured at 96 and 162 MeV
by the Uppsala neutron group.  Furthermore, we have
replaced the original Chew
method by a Difference Method for which the extrapolation is required only
for the difference between the actual cross section and 
that of  a
model with a known value for the coupling constant. The extrapolation now
only concerns a correction and can be done with far greater simplicity and
confidence.  Figure 1 demonstrates concretely how we make such an
extrapolation.  Note the strong improvement in the quality of the
experimental data 
from the older Bonner \cite{BON78}  data to the new Uppsala data 
at the same energy.
\begin{figure}[htbp] 
\begin{center} 
\epsfig{figure=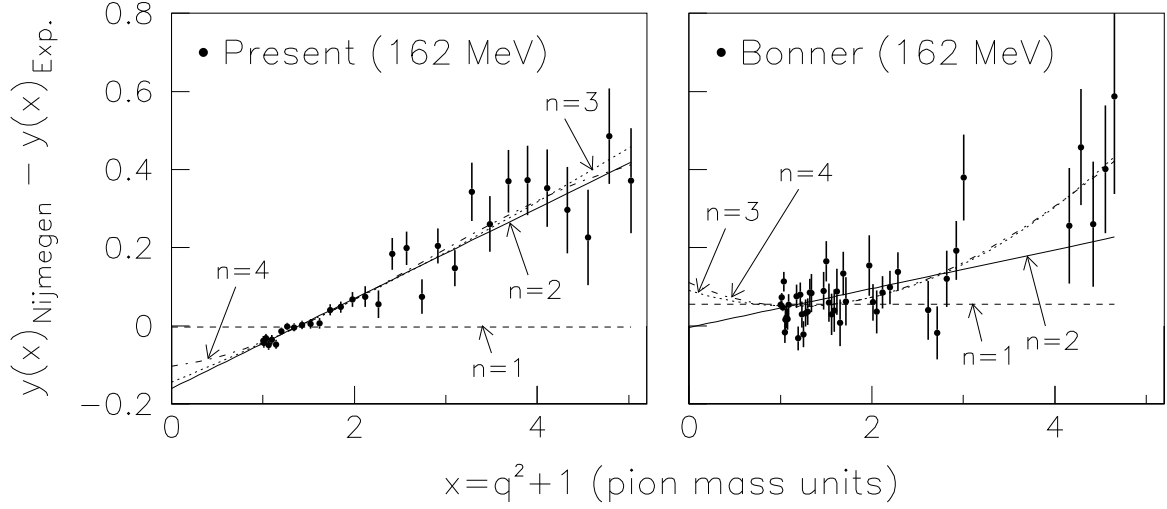 ,width=16cm} \end{center}
\vspace{-1cm}
\caption{Extrapolations  of
the Chew
function $y(q^2)$ to the pion pole at 162 MeV with the Difference Method
using Nijmegen 93  as comparison model and  different polynomial order n.
Left
figure Uppsala data, right figure Bonner data.  For n=2
$g_c^2$(Uppsala)=14.52(26); for n=3 $g_c^2$(Bonner)=12.95(37); the Bonner
data
are
normalized to SM95.} 
\label{fig:diff} 
\end{figure}
How good is this method?  We
have tested it using over 10000 pseudoexperiments generated from models
with known coupling constant with 'experimental' points equivalent to
actual observed ones.  The original coupling constants are regenerated
with an accuracy of about $\pm 1\% $.  The method is therefore well
under control.

The experimental differential cross sections have closely similar shape 
over a wide band of
energies and any energy is as good as another for extrapolation
purposes.  The experimental data from Uppsala have been obtained in dedicated
measurements, in contrast with previous data.  They agree accurately with the
shape of similar experiments at other energies
by the PSI group \cite{Hur80}, 
but differ in shape with data, mainly from Los
Alamos \cite{BON78}. This discrepancy
is presently not fully resolved. (For a different opinion on the Uppsala
data and the extrapolation procedure, see the Comment by  de Swart et
al. and our rebuttal, in  Phys. Rev. Letters {\bf 81} issue 22,
November 30, 1998). A critical discussion of the experimental
situation has been made by Blomgren et al. \cite {Blo}.   Using the most
recent Uppsala data gives $g_c^2= 14.52(26)  $ \cite {Rah98}. 

\section {THE GMO RELATION}

 In order to obtain additional
model-independent
information we (T. E. O. Ericson, B.~Loiseau,  A. W. Thomas) evaluate at
present the Goldberger-Miyazawa-Oehme (GMO) sumrule for $\pi $N forward
scattering \cite {Gol55} in terms of the $\pi N$ scattering lengths and
total
cross sections. Assuming only charge
symmetry:
 \begin {equation}
 g_c^{2} =
-4.50 J^- +103.3 (\frac {a_{\pi ^- p}-a_{\pi ^+ p}}{2}).
\end{equation}
Here $J^-$  is given in $mb$ by the integral
$J^-=-(1/4\pi ^2) \int _{0}^{\infty} (\sigma ^T_{\pi
^+p}-\sigma
^T_{\pi ^-p})/\sqrt {k^2+m_{\pi}^2} dk$  and $a_{\pi ^{\pm} p}$ are expressed
in units of $m_{\pi^+}^{-1}$.

Everything is in principle measurable to good precision. Still this
expression has not been too useful in the past 
because the scattering lengths were theoretically 
constructed
from the analysis of scattering at higher energies.  Recent
splendid experiments at PSI determine the $\pi ^-$p and $\pi
^-$d energy shifts and widths in pionic atoms and from that the
corresponding scattering lengths follow accurately \cite{CHA97}.  We  have
critically examined the
situation with careful
attention to errors.  In particular, we have examined the accuracy of the
constraints due to pion-deuteron data.

 In order to get a robust
evaluation we write the relation as
\begin {equation}
 g_c^{2} =
-4.50J^- +103.3 a_{\pi ^- p}-103.3 (\frac {a_{\pi ^- p}+a_{\pi ^+
p}}{2}).
\end{equation}
Using $J^-=-1.077(47)$ $mb$  \cite {{KOC85},{ARN98}}  and
the experimental $\pi^-$p scattering length \cite {CHA97} 
\begin {equation}
 g_c^{2} =
4.85(22)+9.12(8)-103.3 (\frac {a_{\pi ^- p}+a_{\pi ^+
p}}{2})=13.97(23)-103.3 (\frac {a_{\pi ^- p}+a_{\pi ^+
p}}{2}).\end
{equation}
Here the last term is a small quantity which we
can evaluate with small statistical and
systematic uncertainties from the experimental $\pi ^-$d
scattering length. The cross section
integral $J^-$ is presently the largest source of error. Uncertainties
from the small deuteron term  will not have a
major impact on the result which is stable.
 Evaluating this last term from the impulse approximation only would
increase $g_c^2$
by 1.25(5).  However,  double s wave scattering decreases $g_c^2$ by
$-1.08$, while smaller correction terms come from the p wave Fermi
motion (+0.24), the dispersive correction from absorption ($-0.18$(4)) \cite
{AFN74}  and the
s-p wave double scattering interference term ($-0.21$) \cite {BAR97}.
 To exploit  the present experimental precision the dominant double
scattering term
must be  controlled to better than 10\%, while other corrections require
little more than estimates.  Of these terms the s-p interference term is
presently not fully elucidated.  It depends on short range behavior and
may be partly spurious. Using the correction terms from refs. \cite
{AFN74} and \cite {BAR97} we find a preliminary value
$g_c^2=13.99(24)$ including the s-p interference term and 14.20(24)
excluding it.

In conclusion, we have now two independent methods with controllable
errors for the coupling constant.  The Difference Method gives  14.52(26)
or a 2\% error.  Its future expected improvements are a) a full angular
range, which will give normalization to $\pm 1$\% (now $\pm 2$\%) and b)
several incident neutron energies (which in principle should contain very
similar
information) from which the future precision is expected to reach 
$\pm 1.5$\%.   The GMO relation gives the preliminary value 13.99(24) or
$\pm 2$\%.  The expected improvements are in the dispersion integral
evaluation, now $\pm 4.6$\% to $\pm 2$ to 3\%, which leads to a precision in
the coupling constant of $\pm$(1 to 1.5)\%.

In summary, the two model independent methods which have been critically
examined here provide no support for the low value for the coupling
constant, close to 13.5, which has been advocated elsewhere. 
The lower value cannot be
completely excluded at present, but better data and careful analysis
should settle the issue.

\end{document}